\documentclass[12pt]{article}
\usepackage{graphicx}
\usepackage{amsmath}

\newcommand{\ket}[1]{| #1 \rangle}

\begin{document}

\date{May 2019}

\title{Quantum Reality, Perspectivalism and Covariance.}

\author{Dennis Dieks  \bigskip \\ 
              History and Philosophy of Science, Utrecht University\\Princetonplein 5, 3584CC Utrecht, The Netherlands\\d.dieks@uu.nl}

\maketitle

\begin{abstract}
Paul Busch has emphasized on various occasions the importance for physics of going beyond a merely instrumentalist view of quantum mechanics. Even if we cannot be sure that any particular realist interpretation describes the world as it actually is, the investigation of possible realist interpretations helps us to develop new physical ideas and better intuitions about the nature of physical objects at the micro level.  In this spirit, Paul Busch himself pioneered the concept of ``unsharp quantum reality'', according to which there is an objective non-classical indeterminacy---a lack of sharpness---in the properties of individual quantum systems. 

We concur with Busch's motivation for investigating realist interpretations of quantum mechanics and with his willingness to move away from classical intuitions. In this article we try to take some further steps on this road. In particular, we pay attention to a number of \textit{prima facie} implausible and counter-intuitive aspects of realist interpretations of unitary quantum mechanics. We shall argue that from a realist viewpoint, quantum contextuality naturally leads to ``perspectivalism'' with respect to properties of spatially extended quantum systems, and that this perspectivalism is important for making relativistic covariance possible.
\end{abstract}

\section{Introduction}
\label{intro}
I met Paul Busch for the first time during the 1985 Symposium on the Foundations of Modern Physics in Joensuu, Finland. On that occasion Paul gave a talk on ``Elements of unsharp reality in the EPR experiment'' \cite{busch2}, in which he explained ideas of a research program on which he had embarked not long before \cite{busch1} and to which he would keep returning in his later work.

The Joensuu paper starts from the observation that in any measurement, no matter how carefully designed, there is a non-vanishing residue of uncertainty. In the case of spin measurements, it is thus impossible to construct a spin measuring device that determines spin in exactly one spatial direction. This means that the standard formal treatment of quantum measurements using projection operators should be seen as an idealization, and that actual laboratory measurements are better described with the help of ``smeared out'' projections. In the case of spin measurements this leads to the replacement of projection operators $P_{\vec{n}}$ (for spin in the exact direction $\vec{n}$) by positive operators $E$ that are weighted means  of projections: $ E(\vec{n}) = \int_\Omega d\mu(\vec{n^\prime}) P_{n^\prime}$, with $\mu$ some measure (probability) centered around the direction $\vec{n}$.  More generally, taking into account the approximate nature of measurements results in a variation on the standard quantum mechanical measurement formalism in which  positive operators $E$, \emph{effects}, take the place of projection operators.

The novel interpretative step taken by Paul Busch is to view these effects $E$ not merely as mathematical tools for taking into account the imprecision of measurements on a quantum system, but also as a representation of \emph{unsharp properties} of this  quantum system itself. As he comments \cite[p.\ 351]{busch2} ``It is important to note that `unsharpness' admits not only an interpretation as (subjective) measurement uncertainty but also as proper quantum mechanical uncertainty.'' 

As it turns out, if the unsharpness of the quantum spin effects $E(\vec{n})$ and $E(\vec{n^\prime})$ increases, at a certain point these two unsharp properties will start to admit a joint probability distribution (in spite of the fact that the projection operators $P_{\vec{n}}$ and $P_{\vec{n}^\prime}$ do not commute). It is therefore to be expected that unsharp quantum properties play a role in making the classical limit of quantum mechanics intuitively understandable: unsharpness washes out the consequences of complementarity. Unsharpness of quantum properties also plays an explanatory role in other contexts. For example, the unsharp position of quantum particles makes their behavior in a double-slit experiment fathomable.

Paul Busch fleshed out his realist interpretation of quantum indeterminacy in various later publications \cite{busch3,busch4,busch5,busch6}. In the Introduction of his  review paper with G. Jaeger, ``Unsharp Quantum Reality" \cite{busch7}, the interesting motivation of the authors for pursuing any realist research program at all is made explicit:
 \begin{quote}
 [In physical practice] there is still a strong presence of the view that Quantum Mechanics is nothing more than a formalism for the calculation of measurement statistics... 

It seems to us that a more coherent and productive approach would be to investigate systematically all possible variants of realist interpretations of Quantum Mechanics, including those in which probabilities are not essentially epistemic. On a realist interpretation of Quantum Mechanics as a complete theory, the referent of quantum mechanical propositions is the individual system. This would not only recognize the possibility that such a philosophically realist interpretation could in the end enable the best description of the physical world; it also has the potential benefit of providing us with guidance in developing new, appropriately adapted intuitions about microphysical objects... A primary role of any realist interpretation is to provide a rule that determines, for every state, which physical quantities have definite values in that state, thus representing ``elements of reality'' or actual properties pertaining to the quantum system under investigation. 

In the present paper we put forth an interpretational point of view that has not yet been much considered but that seems to us worthy of further exploration---the concept of \emph{unsharp quantum reality}.
 \end{quote}
 
We fully agree with the motivation for investigating realist interpretations expressed in this quotation. Even though it may be unlikely that we will ever be able to single out one such interpretation as uniquely and faithfully describing physical reality as it is in itself,  each consistent and empirically viable realist interpretation will provide us with a picture of what the physical world \emph{could} be like and will thus enrich our conceptual repertoire and broaden our intuitions.  As stated by Busch and Jaeger, in order to go beyond mere measurement statistics such interpretations of quantum mechanics have to be about physical systems and their properties. 
 
The essential question that arises for each realist interpretation is how the states ascribed  to physical systems by quantum mechanics relate to properties possessed by these systems; different interpretations give different answers to this question. Busch and Jaeger also comment on this issue, and argue that in the case of sharp properties the most plausible answer is provided by the eigenvalue-eigenstate link\footnote{It should be noted that Busch's and Jaeger's defense of the eigenvalue-eigenstate link is independent of their unsharp quantum reality proposal, as emphasized by an anonymous reviewer.}. They reason as follows:
 \begin{quote}
If a property is absent, the system’s action or behavior will be different from that when it is present. Applied to the context of a measurement, in which an observer induces an
interaction between the system and part of its environment (a measurement apparatus), this means that if a property is actual---that is, an observable has a definite value---then its measurement exhibits this value or property unambiguously and (hence) with certainty.

This condition---which, incidentally, is routinely being used as a calibration condition for measuring instruments---is taken as the defining requirement for a measurement scheme to qualify as a measurement of a given observable. Its implementation within the quantum theory of measurement is possible if a property’s being actual is associated with the system being in a corresponding eigenstate. 

In summary, the structure of the quantum theory of measurement ... suggests the adoption of the eigenvalue–eigenstate link as a necessary and sufficient criterion of empirical reality in quantum mechanics.\footnote{Busch and Jaeger immediately add: ``However, the application of quantum mechanics to the description of measurement processes also leads directly into the quantum measurement problem... This problem is one of the main reasons for the continued debate about the interpretation of quantum mechanics.''}
  \end{quote} 
  
The core idea of this argument is that if a perfect measurement of an observable is made on a system possessing  a sharp value of that observable, an outcome revealing this value must be found with certainty, that is with probability $1$. According to quantum mechanics this is only possible if the system is in an eigenstate of the measured observable. This leads to the conclusion that a system can only possess a sharp property (corresponding to an eigenvalue of an observable) if its state is the associated eigenstate---this is the eigenvalue-eigenstate link\footnote{The degree of sharpness (or lack thereof) of unsharp properties can then be characterized via the distance between a system's state and the eigenstate corresponding to the sharp property.}.
  
However, there is room for alternative views concerning this issue: as shown by non-collapse interpretations (e.g., relative-states/many-worlds interpretations \cite{everett,wallace}, modal interpretations \cite{dieksvermaas,lombardidieks}, including the Bohm interpretation \cite{bub}) it is possible to consistently interpret quantum mechanics in terms of physical properties of individual systems even without subscribing to the eigenvalue-eigenstate link. The key idea here is that it is a conceptual possibility that the quantum state specifies epistemic probabilities for the presence of  a range of possible values of a physical quantity, even though only one value is actually realized in any individual case (in the many-worlds interpretation this description applies to the epistemic situation as seen from one single branch). In spite of such a range of non-vanishing epistemic probabilities, a perfect measurement may still unambiguously and faithfully reveal the actually present property with (conditional) probability $1$. In other words, even if the quantum state is not an eigenstate of an observable, so that it specifies non-vanishing probabilities for several eigenvalues, it is not \textit{a priori} inconsistent to assume that an individual system is associated with only one of these values and that this value will be revealed with certainty in a perfect measurement. Of course, if the quantum state assigned to the object is not an eigenstate of the measured observable, this state only allows us to predict probabilities for a range of possible measurement outcomes, and in repetitions of the measurement in this same state these probabilities will materialize as approximate relative frequencies. But nevertheless the principle that a preexisting property of a system must be faithfully and certainly reproduced by an ideal measurement can be upheld if the \textit{conditional probability} for the measuring device to indicate value $a$, if $a$ represents a property of the object system, is $1$.  This shows that the ``calibration condition'' by itself does not prove that a property can only be actual if the system is described by the corresponding eigenstate.

Moreover, as Busch and Jaeger note in their review paper, adopting the eigenvalue-eigenstate link has the unpleasant consequence that we have to accept the occurrence of collapses of the wave function during measurements. Indeed, once a measurement result has become definite, the eigenvalue-eigenstate link tells us that the system on which the measurement has taken place must have ended up in an eigenstate of the measured observable, even if it was in a superposition of such eigenstates before. This leads to the infamous measurement problem: taking collapses seriously entails recognizing restrictions on the validity of  unitary Schr\"{o}dinger evolution, which raises a plethora of well-known problems.  

We shall therefore in the following explore an alternate route, namely acknowledging the universality of unitary evolution and rejecting the eigenvalue-eigenstate link as a necessary condition for the attribution of (sharp) properties\footnote{The eigenvalue-eigenstate link as a \emph{sufficient} condition for property attribution is unproblematic. \emph{Which} properties are to be ascribed to a system when the eigenvalue-eigenstate link is abandoned, depends on the specific unitary interpretation that is considered.}.  As we shall see, attributing properties to systems in the context of unitary quantum mechanics leads to an unexpected, intuitively peculiar non-classical picture.

\section{Unitary Quantum Mechanics}
\label{unitary}

There are several interpretational schemes only admitting unitary evolution and excluding collapses: the Bohm theory, consistent histories, Everett/many worlds, and modal interpretations are examples. These schemes all start from the basic view that the same physical principles should apply regardless of whether we are dealing with elemental physical interactions or complicated measurements.  Measurements are accordingly treated as ordinary physical interactions, between object systems and measuring devices. Given the huge amount of evidence that outside the measurement context physical interactions between quantum systems  are governed by unitary evolution (which leads to superpositions and entanglement), the assumption that unitary evolution applies universally becomes natural once one does not assign an independent status to measurement interactions.

Unitary schemes not only unify ordinary physical interactions and measurements, but also the theoretical treatment of the micro and macro levels. Macroscopic properties like the position of a pointer on the dial of a measuring device should accordingly fall under the same rules of quantum property ascription as  microscopic properties. Even human observers should in principle be seen as very complicated quantum systems,  analyzable in terms of the physical properties they possess (including memory contents). This general programmatic viewpoint  is of course not unique to quantum physics: it generalizes physicalist ideas well known from  classical physics. We mention the point here because it will play a role in our later discussion of measurements (and their reversal) in unitary quantum mechanics.

As discussed in the previous section, the argument that a system can only possess a determinate value of a physical quantity when it is in the appropriate eigenstate of the associated observable can be circumvented by a probabilistic interpretation of the quantum state, according to which this state provides us with probabilities for individual eigenvalues to be present. Unitary schemes employ the usual Born probability rule for these probabilities\footnote{Often these Born probabilities are simply posited; some authors attempt to derive them from more basic principles, e.g.\ \cite{wallace}. For our purposes this difference is not important.}. 

A point that will play an important role later on is that the assumption of universal validity of unitary Schr\"{o}dinger evolution entails that all quantum processes can in principle be reversed. Just as in classical physics any process that goes from initial state $A$ to final state $B$ can be undone by a physical interaction in the reverse direction, leading back to the initial state, the final state in a quantum process governed by unitary interaction can also be brought back to its initial state by applying appropriate interactions. In particular, it is possible (in principle!) to undo a measurement: after a record of the measurement result has been formed, an appropriate reversal of the interactions between the elementary constituents involved in the process will be able to restore the situation that was present before the measurement---the record of the outcome will of course be erased during this process.

Of course, a restoration of the initial situation is also possible in schemes in which collapses occur: after a projection that takes $\ket{\Psi} = \Sigma_i c_i \ket{\psi}_i$ to $\ket{\psi}_k$, there will in principle always be a unitary evolution that transforms $\ket{\psi_k}$ into $\ket{\Psi}$ again, since unitary transformations are rotations in Hilbert space and any vector can be rotated into any other.  The essential difference with the unitary case becomes clear when we consider states of several systems that are correlated, as in the EPR situation. For example, after a collapse $\Sigma_i c_i \ket{\psi}_i^1 \otimes \ket{\varphi}_i^2 \Rightarrow \ket{\psi}_k^1 \otimes \ket{\varphi}_k^2$, as a result of a measurement on system $1$, the initial entangled state  can only be reproduced via a transformation that affects the total system, comprising both $1$ and $2$. By contrast, when we assume only unitary evolution an ideal measurement on system $1$ alone is represented by a transition of the form \[\Sigma_i c_i \ket{\psi}_i^1 \otimes \ket{\varphi}_i^2 \otimes \ket{E}_0 \Rightarrow \Sigma_i c_i \ket{\psi}_i^1 \otimes \ket{\varphi}_i^2  \otimes \ket{E}_i ,    \] in which $\ket{E}_i$ and $\ket{E}_0$ are states of a measuring device interacting solely with system $1$. This transition can be reversed by locally acting on system $1$ and its measuring device; no interaction with system $2$ is needed.  For the case of EPR experiments on space-like separated systems this implies that the effects of a local measurement on system $1$ can also be undone locally, by interactions with $1$: the total state, including system $2$, can be brought back to what it initially was without touching system $2$ (something that is impossible in the case of collapses).     

\section{Unitarity and Relativity}
\label{gao}

Unitary interpretations as defined in the preceding section are known to face problems with Lorentz invariance (e.g., \cite{berndl,dickson,myrvold}). The nature of the difficulty  has recently been illustrated in a simple way by Gao \cite{gao1}. In Gao's thought experiment two observers, Alice and Bob, are located at space-like separation from each other and perform spin measurements on an EPR pair of electrons in the singlet state (the usual Bohm-EPR setup). Let us assume that Alice and Bob both measure spin in the same direction, so that quantum mechanics predicts that their outcomes will be anti-correlated. 

As we have seen a moment ago, unitary quantum mechanics permits the local reversal of such measurements, which restores the original two-particle singlet state. This can be used to construct a paradoxical variation on the usual  Bohm-EPR thought experiment. 
 
Let Alice measure the spin of her particle and let her record (or memorize) the outcome---we may think of Alice as fulfilling the role of the measuring device in the discussion at the end of section 2. After Alice's measurement a unitary restoration process takes place that locally undoes the measurement and returns Alice and Bob plus the two particles to their original total state (so that Alice loses all her information about the outcome). Finally, Bob measures the spin of his particle in the same direction in which Alice had measured before.
 
Let us call the inertial frame in which the just-given description of the experiment applies the laboratory frame. When we apply the standard quantum mechanical rules in this frame, we predict that Alice records either $+1$ or $-1$ as her result, both possibilities having probability $1/2$. Alice's record is erased in the subsequent anti-measurement, after which Bob performs his measurement in the original singlet state so that he also finds either $+1$ or $-1$, again with probability $1/2$ for each possible result.   

However, because Alice and Bob are space-like separated, there exists another inertial frame of reference with a simultaneity relation such that in this frame Alice  measures her electron first, then Bob measures his electron, and finally Alice's measurement is undone only \emph{after} both measurements have taken place.

This leads to the following problem. In the second frame, in which Alice's experiment is undone as the final step, Bob's measurement \emph{must} yield $-1$ if Alice's result was $+1$, because of the anti-correlation in the singlet state: the conditional probability of Bob's outcome being $-1$ is $1$, given Alice's result. As we have seen, however, in the laboratory frame Alice's result no longer exists when Bob performs his measurement, since Alice's measurement has been undone at that instant and the original Bohm-EPR state has been restored. Consequently, according to the usual quantum mechanical rules applied in the laboratory frame Bob's result does not need to be $-1$ even if Alice's result was $+1$; Bob may alternatively find $-1$, with probability $1/2$. 

The consequence is that in a long series of repetitions of this ``EPR plus restoration'' experiment, after selection\footnote{That is, selection \textit{in abstracto}, with the mind's eye. Any real physical selection would involve an interaction that would spoil the perfect reversal of Alice's measurement.} of all cases in which Alice's experiment yields $+1$, we have to expect that Bob finds $-1$ and $+1$ in approximately half of the cases according to quantum mechanics applied in the laboratory frame, but always $-1$ according to quantum mechanics in the other frame. 

We may construct a variation on this thought experiment in which we do not need a series of repetitions with new EPR pairs, but use the same electrons over and over again. Suppose that when Alice's experiment has been undone (as seen from the laboratory frame), and Bob has completed his measurement, Bob's measurement is undone as well, after which he performs a second measurement. This second measurement can be reversed in its turn, after which a third measurement can be done; and so on. If the space-time distance between Alice and Bob and the time scales of Bob's measurements are judiciously chosen  the experiment can be arranged in such a way that all Bob's measurement take place before Alice's measurement is undone, according to the simultaneity in some other inertial frame. In this way we arrive at the same paradox as before: In the lab frame quantum mechanics tells us that approximately half of Bob's results will be $-1$ if Alice's result was $+1$, in the other frame the quantum prediction is that all of Bob's outcomes must be the same, namely $-1$.   

In both cases we arrive at a contradiction, which suggests that the unitary formalism of quantum mechanics cannot be used for making predictions in the same way in all inertial frames. 

One way of responding to this is to posit that there exists a preferred inertial frame, whose simultaneity is the only one to be used when predicting outcomes and their frequencies with the standard formalism of quantum mechanics. This is the route taken by the Bohm theory---that the Bohm theory faces problems with special relativity and cannot accept the equivalence of all inertial reference frames has long been recognized because of the occurrence of instantaneous interactions in that theory. It is interesting to discuss in some more detail how the Bohm theory deals with the new thought experiment, and to compare it with other unitary schemes.

According to the Bohm theory the complete state of a system of two spin-$\frac{1}{2}$ particles is given by its wave function \emph{plus} the individual positions of the two particles. We consider here the standard version of the theory according to  which the particles do not possess an intrinsic spin---their only physical property is position. However,  the wave function is taken to be the usual spinor-valued one (in our case with the singlet state as its spin part) and the Hamiltonian governing the evolution of the wave function contains the usual coupling between spin variables and magnetic fields.   In the interaction with a Stern-Gerlach device the spatial wave function will therefore be split into two parts, an upper and a lower wave packet. The initial positions of the particles, in their initial wave packets, determine whether they will end up in the upper or the lower wave packet on their wing of the experiment. In this way the final position of a particle, after the interaction with a Stern-Gerlach device, represents the result of a ``spin measurement'' even though the Bohm theory does not operate with an intrinsic spin property. The result of the spin measurement is either ``up'' or ``down''.  Because of the non-local interaction between the particles (due to the entanglement of  the total state), particle $2$ will be driven down when particle $1$ is the first to interact with the magnetic field of the Stern-Gerlach device and goes up; and \textit{vice versa} when particle $2$ is the first to interact (see \cite{norsen} for a detailed discussion).   

In the initial stage of our thought experiment there is one spatial wave packet near Alice and one near Bob. Let us suppose that both Alice's and Bob's particle find themselves in initial positions such that they will emerge ``up'' (spin result $+1$) when they are the first to interact with their respective Stern-Gerlach devices (the particle that is  second to interact will go opposite to the other one, so that its spin outcome will be $-1$ and the anti-correlation between the spin results predicted by quantum mechanics is reproduced). In repetitions of the experiment with new pairs of electrons there will be a probability of $1/2$ for each electron to be in an initial position, in its own wave packet, that leads to the result ``up''\footnote{From this description one already sees why the notion of a preferred frame imposes itself in the Bohm theory: if Alice and Bob are at  space-like separation from each other, the time order of their spin measurements will generally be frame-dependent. In this case, application of quantum mechanics in the same way in all frames leads to contradictory predictions for the measurement outcomes.}. 

Suppose that as judged from the laboratory frame Alice makes her measurement first and finds $+1$, after which a reverse local evolution takes place that restores the initial wave function, returns Alice's particle to its original position, and erases Alice's result. Then Bob measures the spin of his particle and also finds $+1$, the outcome deterministically determined by his particle's  initial position. In a long series of repetitions of the experiment, each time with a new pair of electrons, and collecting (in thought!) all runs in which Alice's outcome is $+1$, the Bohm prediction is that Bob will find both instances of $+1$ and of $-1$, each of these outcomes in approximately  $50 \% $ of the cases---in the Bohm theory it is supposed that in repetitions of the experiment the initial particle positions will be distributed according to the Born rule.  By contrast, if Bob's measurement is repeated many times with the same particle, each time undoing the measurement and restarting from the same initial situation, the Bohm prediction is that the same result ($+1$) will be found in each run of the experiment.

When we now consider these three versions of the experiment from the other inertial frame, in which Bob's measurements take place before Alice's measurement is undone, we find the following.  In the single-run experiment Alice finds $+1$ and Bob $-1$, because the non-local influence of Alice's result sends Bob's particle ``down''. When the experiment is repeated many times with new particle pairs, and Alice's $+1$ cases are collected, Bob will in all cases measure $-1$. If the experiment is done repeatedly with one particle pair, as described before, Bob will also find $-1$ each time.

Clearly, the predictions from the two frames of reference contradict each other in all three versions of the experiment. This further supports  the necessity of a privileged frame in Bohm's theory.

Let us now consider the same experiments from the perspective of non-Bohmian unitary quantum mechanics (i.e., unitary schemes that do not assume that position is a preferred observable \cite{bub}). According to the standard quantum rules applied in the lab frame, Alice will find either $+1$ or $-1$, both with probability $1/2$. Let us again focus on the case with outcome $1$, in order to compare with the predictions of the Bohm theory. After the reversal of Alice's measurement Bob finds either $+1$ or $-1$, with equal probabilities. In a long series of experiments, with many pairs of particles, we find that Bob's results are $50\%$ up and $50\%$ down. If Bob's experiment is repeated by means of reversals we also find a $50\%-50\%$ distribution.   

There is therefore a difference with the Bohm prediction for the third version of the experiment. Because of its determinism, the Bohm scheme predicts that repetitions of an experiment with exactly the same initial conditions will lead to exactly the same outcomes. But since the initial conditions in non-Bohmian quantum mechanics do not involve particle positions, this Bohm result cannot be carried over to the standard theory. So it might seem that we have a situation where we could, in principle, empirically distinguish between the Bohm theory and standard quantum mechanics. This is deceptive, however: because Bob's measurements are each time undone, no records of them can exist and no empirical comparison of their outcomes is possible. Whether or not there is a correlation between the results of repeated measurements on the same system, with reversals between the measurements,  is something that as a matter of principle cannot be verified or refuted empirically.

When the second frame of reference is used to describe the three versions of the EPR-plus-reversal experiment with non-Bohmian unitary quantum mechanics, we find that Bob measures $-1$ if Alice's result is $+1$. In repetitions with new particle pairs Bob will also always find $-1$ in the cases in which Alice measured $+1$. Finally, when Bob's measurements are repeated on the same particle he will still find $-1$ in all cases. 

Non-Bohmian quantum mechanics, applied in the second frame, thus yields the same predictions as Bohm's theory: the $+1$ result of Alice fixes Bob's outcome as $-1$ in all cases. Again, the predictions from frame $2$ run counter to what is predicted from the laboratory frame. In the first version of the experiment (one single measurement by both Alice and Bob) the probability of Bob finding $-1$ is $1/2$ in the lab frame instead of $1$ (the latter probability value is predicted in the other frame).  Although this is not an direct contradiction in terms of measurement results, it becomes so in the second and third versions of the experiment in which the probability is made into an approximate frequency.

Given that the Bohm theory has already developed a tool to protect itself against relativistic inconsistencies, namely the postulation of a privileged reference frame with respect to which the non-local quantum interactions are defined, it seems only natural to take over this recipe for unitary interpretations in general. It is true that assuming the existence of such a frame is at odds with Einstein's first special relativistic postulate; but it has often been argued that this objection is not decisive because quantum mechanics contains an element of non-locality and therefore is at odds with special  relativity anyway. 

However, it is possible to interpret this non-locality of quantum mechanics in a way that does not involve superluminal causation or action-at-a-distance, and does not conflict with relativity. A key to this alternative viewpoint is the observation that physical systems as described by quantum mechanics are generally not space-time objects in the way classical systems are.  As we shall argue, this observation may be used as a starting point for the construction of a new scheme for reconciling quantum mechanics, and the above thought experiments, with relativity---even though the ensuing alternative solution of the inconsistency problems will require a break with a number of classical intuitions. 

\section{Perspectives and Non-Locality}
\label{perspectives}

According to classical physics physical systems can be analyzed as built up from local parts, each part possessing its own locally defined properties. Specification of all these local parts and their properties provides us with the complete instantaneous state of a classical system as a whole. For example,  a measuring rod is described as a collection of very many particles whose instantaneous local properties together determine the properties of the composite system. A classical field in a spatial region is likewise completely specified by the local field strengths at all points in that region. 

The formalism of quantum mechanics suggests a different picture, though.  This is because in general many-particle states are entangled, and such entangled states are not combinations of one-particle states that specify complete sets of one-particle properties. The Bohm-EPR two-electron singlet state furnishes a typical example.  This state is the product of a symmetric spatial part and an anti-symmetric spin part 
 \begin{equation}\label{EPReq}
| \Phi \rangle = \frac{1}{\sqrt{2}} \{| L \rangle_1 | R
\rangle_2 + | R \rangle_1 | L \rangle_2 \} \otimes \{ |\!\uparrow
\rangle_1 |\downarrow \rangle_2 - |\!\downarrow \rangle_1
|\uparrow \rangle_2 \},
\end{equation}
where $| L \rangle$ and $| R \rangle$ correspond to narrow wave packets localized on the left and right wing of the EPR experiment, respectively. If one were to think of the EPR experiment as pertaining to two particles, each one with its own complete set of spatial and spin properties, one would expect the total state to be built up from the one-electron states $\ket{L}_1\ket{\!\uparrow}_1, \ket{L}_1\ket{\!\downarrow}_1, \ket{R}_2\ket{\!\uparrow}_2 $ and $ \ket{R}_2\ket{\!\downarrow}_2 $, in each of which a full set of particle properties is specified (namely a definite spin plus a localization property). Instead, however, we find that in Eq.(\ref{EPReq}) the global spatial features of the two-particle system and its global spin properties are independently combined, i.e.\ without a correlation between individual particle localization and individual spin. 

In fact, the indices $1$ and $2$, usually taken to be particle labels, are not correlated to either $\ket{L}$ or $\ket{R}$ so that in state (\ref{EPReq}) we cannot even speak about particle $1$ as being at the left-hand side and particle $2$ at the right-hand side, or \textit{vice versa}.  As argued in  \cite{diekslubb1,dieksinfo,diekslubb2} this shows that the standard interpretation of state (\ref{EPReq}) in physical practice, namely as representing two particles at a large distance from each other, one on the left and one on the right, implicitly renounces the doctrine that the indices $1$ and $2$ label these particles. Rather, in physical practice the individual constituents of the total system are taken to correspond directly to the two spatial one-particle states  $\ket{L}$ and $\ket{R}$ ($L$ referring to the left particle and $R$ to the right one). However, even with this understanding there is no correlation between spatial properties and definite spin directions, so that we still cannot think in terms of a particle on the left possessing its own definite spin and a similar particle on the right. Instead, Eq.(\ref{EPReq}) represents a system with a global spin property that is uncorrelated to its localization properties.  

In other words, the classical paradigm according to which a system consists of independently defined local parts does not fit in in a natural way with how quantum mechanics generally describes physical systems. As argued in \cite{diekslubb2,diekspersp}, violations of Bell inequalities can accordingly be viewed as evidence that a classical ``local parts picture'' does not apply---instead of the more common interpretation that such violations are manifestations of non-local interactions or influences between distant parts.

In order to fix the properties of a two-particle system in an entangled quantum state like (\ref{EPReq}) we therefore need to specify its total, global state, which cannot be written as a concatenation of local states or as the combination of two one-particle states. In the case of a spatially extended system as in (\ref{EPReq}) the specification of the total state must involve the identification of a hyperplane of simultaneity, or more generally a space-like hypersurface, on which the state is defined. Different choices of such hyperplanes will generally lead to different property attributions (cf. \cite{dieks1}).  The thought experiment of section \ref{gao} provides us with an example: the laboratory frame and the other inertial frame considered there  define  alternative hyperplanes of simultaneity on which the global states of the system are different. On one hyperplane the total state is correlated to a state of Alice's measuring device corresponding to a definite outcome, whereas there is no such correlation on other hyperplanes. It cannot be excluded \textit{a priori} that this leads to differences in the description even locally at points where  the two hyperplanes intersect, in particular at Bob's position.  

What we accordingly propose as an interpretative option is that the physical characteristics of what happens in Bob's measurements not only depend on the local circumstances near Bob, but also on the total state that determines the properties of the system that is being measured and thus on the hyperplane on which that state is defined. If the global state is different on different hyperplanes, the properties of the global system will generally be different as well so that Bob as described on different hyperplanes will be involved in measurements of different properties. 

What we are suggesting here is the attribution of hyperplane dependent, and in this sense  \textit{perspectival} properties. The idea of perspectivalism is not new: see for example \cite{rovelli,benedieks,dieksrel,laudisa}. These earlier proposals discuss property attributions to a system from different perspectives connected to different observers (who stand in different relations to the system depending on whether or not they have interacted with it).  The present proposal is even more radical because it contemplates the possibility that the same observer measures different outcomes depending on the hyperplane on which the measurement is described (see, however, \cite{fleming} for arguments that hyperplane dependence is not as radical as it seems). 

A comment that may come to mind is that hyperplane dependence is something already well known from special relativity. However, although it is true that in special relativity different frames of reference are associated with different descriptions of physical processes, these descriptions relate to each other via Lorentz transformations that translate between states of affairs defined point by point. These Lorentz transformations will never translate an outcome $+1$ (the event of a pointer indicating $+1$ on a dial)  in one frame of reference into the outcome $-1$ in another frame.  
The hyperplane dependence that we are discussing here goes further than this special relativistic frame dependence. The non-quantum special relativistic picture is still based on the principle that physical systems and processes are built up from independently existing locally defined events.  But in the quantum mechanical formalism of many-particle systems physical properties are not specified per point. The resulting hyperplane dependence is different from what is at issue in special relativity with its point-by-point Lorentz transformations\footnote{That many-component systems are defined globally rather than point by point  is also important for the covariance properties of wave function collapse construed as an effective description within unitary quantum mechanics \cite{dieks1}.}.

\section{Uniqueness of Outcomes}
\label{uniqueness}

According to this perspectivalist proposal there may be a difference between Bob's outcome in relation to Alice having made a not yet undone measurement and with respect to Alice after her measurement has been undone. Evidently, this conflicts with the intuition that it cannot make any difference for Bob's local situation what happens far away---the locality notion that underlies relativity and the EPR argument. As pointed out before, this intuition depends on the idea that any global system consists of spatial parts, each with its independently defined properties---a notion of which we have seen that it does not sit well with the mathematical structure of quantum mechanics. 

But, it may be objected, perhaps Bob is not aware that Alice is performing measurements far away, or does not even know that the system on which he is making his measurement has a distant counterpart. Surely, Bob will find one definite result completely independently of such considerations? All our experience indicates that any successful experiment has only one definite result. Can this be reconciled with the idea that Bob's results are perspectival?

The uniqueness of direct experience is not in conflict with perspectivalism, because in each perspective there is only one unique and definite measurement result.  But it is true that there is a multiplicity of outcomes corresponding to different perspectives. This is reminiscent of the many-worlds interpretation of quantum mechanics, in which there are many branches, each with its own unique state of affairs. But although there is this similarity with the many-worlds account, there are also important differences. 

In relativistic many-worlds branching \cite{bacciagaluppi}, Bob's local measurement  induces a local splitting of worlds so that each possible spin plus position outcome is  instantiated in one of the resulting branches. This splitting then propagates through the universe, with a velocity less than or equal to the velocity of light. This account retains the classical notion of locality: the branching events arising from Alice's and Bob's measurements are localized at Alice's and Bob's positions and the branching propagates like a wave through space-time. 

By contrast, in the perspectival account discussed here there is no local propagation of different possibilities (branches) but two different states of affairs as defined on two different global hyperplanes. Moreover, in our account not all possible results need to be instantiated. After one pair of measurements by Alice and Bob, Alice's spin outcome may be $+1$ and Bob's outcome $-1$, from all perspectives; this is a globally consistent possibility. It is true that in repetitions of Bob's measurement  the outcome $+1$ will also occur, namely with respect to hyperplanes on which Alice's measurement has been undone.  This is not the consequence of a splitting of the world, however. It takes place in the same world as before, although along a different hyperplane from the one on which the outcome $-1$ is instantiated.

So there is a multiplicity of outcomes in a quite specific and limited sense, namely relating to different perspectives defined by different hyperplanes, all occurring in one single world. This restricted multiplicity is sufficient to avoid a number of no-go results that may seem to show that quantum experiments cannot have unique outcomes in a stronger sense, and that even the objectivity and consistency of quantum theory are under threat \cite{frauchiger,bub2,healey}.

\section{Perspectives and Contextuality}
\label{contextuality}

A recent example of such no-go arguments was recently discussed by Healey,  who credits its idea to Masanes  \cite{healey}. As in the thought experiment discussed in section \ref{gao}, Alice and Bob (who find themselves at space-like separation from each other) share a Bohm-EPR pair of spin-$\frac{1}{2}$ particles and each of them is set to perform a spin measurement, in directions $\vec{a}$ and $\vec{b}$, respectively. In the new experiment there are also two colleagues of Alice and Bob, Carol and Dan, with their own measuring devices; Carol finds herself close to Alice, Dan is located near to Bob. Carol and Dan are set to make spin measurements in the respective directions $\vec{c}$ and $\vec{d}$.  As in the earlier experiment, it is supposed that measurements can be undone by local operations, as permitted by unitary quantum mechanics; Alice and Bob are assumed to possess the means to locally reverse the measurements made by Carol and Dan, respectively.   

Measurement are now made in the following way and order. First Dan measures the spin of the particle near to him in the $\vec{d}$ direction, then Carol performs a measurement on her particle, in the $\vec{c}$ direction. Subsequently, Alice undoes Carol's measurement and performs her own measurement in the $\vec{a}$ direction. Because Carol's measurement has been undone, the state of the particle system plus Dan is the same for Alice, when she does her measurement, as it was for Carol when she  did hers. Finally, Bob undoes Dan's measurement and measures in the $\vec{b}$ direction himself.

So in the final situation measurements have been made in the four directions $\vec{a}, \vec{b}, \vec{c}$ and $\vec{d}$ on one and the same Bohm-EPR pair. This distinguishes the present thought experiment from standard EPR-type experiments: in the standard cases Alice measures in her direction and Bob in his, and the experiment has to be repeated with other particle pairs  in order to make measurements in other directions. 

In the lab frame in which the four experimenters are at rest the correlations between the outcomes found by Dan and Carol, Alice and Dan, and by Alice and Bob, respectively, can be directly calculated from the standard formula $-\cos (\theta)$, with $\theta$ standing for the angle between the directions in which the two parties involved have measured their spin values.  The correlation between the results of Bob and Carol cannot be calculated this way in the laboratory frame, because Carol's result has already been erased when Bob performs his measurement.  But this complication can be circumvented by a  maneuver that is reminiscent of the earlier thought experiment:  there exists another inertial frame in which Bob, who is at space-like separation from Alice and Carol, makes his measurement before Alice reverses Carol's measurement and performs her own. In this frame, the correlation between the outcomes found by Carol and by Bob can again be calculated by means of  the standard formula $-\cos (\theta)$. 

Now, if we assume that all four measurements have outcomes that are perspective-independent, we have in a single run of the experiment the four jointly existing results $a, b, c$ and $d$. In repetitions of the experiment there will in this case be a probability distribution over these values. This is exactly the familiar situation in which a Bell inequality can be derived: the values $a, b, c$ and $d$ play the role of the local hidden variables that are assumed to exist in the standard Bell derivation. But the correlations between the measurement outcomes, as predicted by quantum mechanics in the way just explained, violate that Bell inequality if the directions $\vec{a}, \vec{b}, \vec{c}$ and $\vec{d}$ are appropriately chosen.  

The assumption that the outcomes of the four measurements in a single run of the above experiment can be represented by unique numbers $a, b, c$ and $d$ that possess a joint probability distribution thus leads to a contradiction\footnote{This contradiction is confirmed in an interesting way by a calculation of Gao's \cite{gao2}, who notes that if a joint distribution of $a, b, c$ and $d$ exists, the joint distribution of $b$ and $c$ is completely fixed by the joint distributions of $c$ and $d$, $a$ and $d$, and  $a$ and $b$, respectively.  Direct calculation, using the quantum predictions for the latter three joint distributions, demonstrates that the thus found joint distribution of $b$ and $c$ does not agree with what quantum mechanics predicts for it.}.  

The inconsistency disappears if we introduce context-dependence: in that case the outcome of Carol's measurement in the context of Dan's result may be different from Carol's outcome in the context of Bob's result. This solution accords with our proposal in section \ref{perspectives}: hyperplanes connecting Bob's and Carol's measurements are different from hyperplanes connecting the measurements by Dan and Carol, and the perspectivalism explained earlier tells us that in this case Carol's result need not be the same on these two hyperplanes. 

The just-discussed thought experiment, in which there is a conflict with Bell inequalities, sheds additional light on the status of quantum perspectivalism. An often-heard comment on violations of Bell inequalities is that they signal the presence of a non-local influence between distant events: different choices of the direction in which to measure a spin value at one position apparently make a difference for values of quantities far-away. But in the variation on the EPR setup with four experimenters and reversals of measurements, there are no choices to be made: all measurement directions are instantiated together and fixed in each run of the experiment. Consequently, an interpretation in terms of superluminal influences due to distant choices is ruled out. The role of such non-local influences in avoiding violations of Bell inequalities is taken over by the perspectival nature of the measurement outcomes: it is not the case that different choices of directions at Bob's and Dan's locations change something at Carol's position. Rather, Carol's physical quantities are  differently defined with respect to Bob's and Dan's results. 

This is basically a case of \textit{contextuality}:  the physical quantity measured by Alice is different in the context of Bob's measurement from what it is in the context of Dan's measurement. It is well known that contextuality of this sort is an essential feature of quantum mechanics, as proved by the Kochen and Specker theorem. This same contextuality can be identified as the basis of the violation of Bell inequalities (cf., e.g., \cite{mermin}). Indeed, the incompatibility in the Bohm-EPR experiment between quantum mechanics and the joint existence of values of the four possible measurement results does not depend at all on the mutual distance between the parts of the system: it persists in exactly the same way when there is no appreciable distance between them. The consideration of great distances between the partial systems only serves to bring out the strange consequences of a ``classical'' interpretation of this contextuality in terms of causal influences: if separations become space-like, such influences have to be superluminal or even instantaneous.  What the perspectivalism proposal does is to replace all such talk about influences, causality and propagation speeds by the idea that a context---or perspective---is needed to \emph{define} the value of a physical quantity, in accordance with the core of the Kochen and Specker results (and, one might add, of certain old Bohrian ideas).

In some more technical detail, the Kochen and Specker theorem demonstrates that if we want to assign a value to a physical quantity of a part of a composite system represented by an operator $A$, we have to take into account the rest of the total system, in the following sense. It makes a difference whether $A$ is considered as a function of  $A\otimes B$ or of $A \otimes C$, where $B$ and $C$ are observables of the rest of the system that do not commute with each other. The Kochen and Specker theorem shows that the assumption that $A$ can be assigned one and the same value in all such combinations with observables of the remainder of the system leads to a contradiction.  This result is based on the structure of the algebra of observables in Hilbert space, and holds independently of any consideration of the distance between the systems to which $A$, and $B, C$, pertain. The perspectivalism that we have discussed is in line with this: physical quantities are generally defined with respect to quantities in other systems, which in the specific case of spatially extended systems may be expressed as hyperplane dependence\footnote{Of course, this perspectivalism or hyperplane dependence becomes significant only in the case of entangled states. If the total state is a product state, or if the effects of entanglement are washed out by a decohering environment, monadic, non-perspectival properties  can be ascribed to the two partial systems without contradiction.}.  
   
\section{Conclusion: Unitary Quantum Mechanics and Covariance}
\label{covariance}

Unitary quantum mechanics faces a difficulty with relativistic covariance if it is interpreted in terms of space-time events that are characterized by locally defined properties: it can be rigorously proved that no unitary scheme that provides such a local account can satisfy the requirement that the same probability rules apply equally to all hyperplanes in Minkowski space-time \cite{berndl,dickson,myrvold}. 

The core premise in these proofs is that intersecting hyperplanes should carry local properties in a coherent way, so that they provide agreeing descriptions of the physical conditions at the space-time points where they intersect. The proofs demonstrate that this local meshing of hyperplanes is impossible, if the Born probability rule is to be valid on all hyperplanes. 

One way of responding to this situation is to introduce a privileged inertial frame of reference. This is the course taken by the Bohm theory, according to which the Born rule only holds in the preferred frame. This is a consistent and empirically viable way out of the problem; but still, there is a tension with the spirit of special relativity. Indeed, Einstein in 1905 made the equivalence of inertial frames a fundamental principle of his theory in order to theoretically express the empirical fact that measurement results do not provide evidence for a privileged role of any particular frame. This motivation for the first postulate keeps its force also in quantum mechanics: the empirical predictions of quantum mechanics, even in the Bohm interpretation, are such that they make any assumed preferred frame  undetectable. 

It is therefore worth-while to ask whether there exist other options for  escaping the no-go theorems.  The answer is affirmative: such a possibility has in fact been outlined in the foregoing sections. Indeed, as we have seen, the crucial assumption in the no-go theorems is that properties of physical systems are monadic, in the sense of independent of the presence of other systems and locally defined independently of the hyperplane on which they are considered. This assumption is rejected in interpretations that work with perspectival and hyperplane dependent properties. In these interpretations physical properties are defined per hyperplane, in a way that is relativistically covariant. 

Unlike the proposal to accept the existence of a privileged inertial system, the idea to accept that physical properties are perspectival is not \textit{ad hoc}, because it addresses more questions than merely the lack of Lorentz invariance and fits in naturally with the mathematical structure of quantum mechanics. In particular, as argued in section \ref{contextuality}, it may be seen as an expression of the contextuality that the Bell, and Kochen and Specker, results have shown to be central to quantum theory.  

Accepting that physical properties are not monadic and locally defined, but rather perspectival, relational and hyperplane dependent is a huge step away from everyday experience and from the intuitions that served us well in non-quantum physics. But it  constitutes a broadening of our conceptual repertoire of exactly the kind that Paul Busch envisaged as a benefit of investigating realist interpretations of quantum mechanics.

\end{document}